\documentclass[fleqn,usenatbib]{mnras}

\usepackage{newtxtext,newtxmath}

\usepackage[T1]{fontenc}

\DeclareRobustCommand{\VAN}[3]{#2}
\let\VANthebibliography\thebibliography
\def\thebibliography{\DeclareRobustCommand{\VAN}[3]{##3}\VANthebibliography}

\usepackage{graphicx}	
\usepackage{amsmath}	
\usepackage{acronym}
\usepackage{xspace}
\usepackage{booktabs}

\acrodef{lpt}[LPT]{long period radio transient}
\acrodef{wd}[WD]{white dwarf}
\acrodef{cv}[CV]{cataclysmic variable}
\acrodef{ip}[IP]{intermediate polar}
\acrodef{snr}[SNR]{signal-to-noise ratio}
\acrodef{sed}[SED]{spectral energy distribution}
\acrodef{rv}[RV]{radial velocity}
\acrodef{larp}[LARP]{low accretion rate polar}
\acrodef{uv}[UV]{ultraviolet}
\acrodefplural{rv}{radial velocities}

\DeclareUnicodeCharacter{2013}{\textendash}

\newcommand{\jf}{ASKAP J1448$-$68\xspace}

\title[ASKAP J1448-68 is a CV]{The long period radio transient ASKAP J1448$-$68 is a cataclysmic variable}

\author[G. Mo et al.]{Geoffrey Mo,$^{1,2}$\thanks{E-mail: gmo@carnegiescience.edu}
Akash Anumarlapudi,$^{3}$
Iris de Ruiter,$^{4,5}$
Kareem El-Badry,$^{2}$
David L. Kaplan,$^{6}$
\newauthor
Antonio C. Rodriguez,$^{7}$
Ilaria Caiazzo,$^{8}$
Kaitlyn Shin,$^{2}$
Anthony L. Piro,$^{1}$
Tim Cunningham,$^{9}$ and
\newauthor
Daniel D. Kelson$^{1}$
\\
$^{1}$The Observatories of the Carnegie Institution for Science, 813 Santa Barbara St., Pasadena, CA 91101, USA\\
$^{2}$Department of Astronomy, California Institute of Technology, 1216 E California Blvd, Pasadena, CA 91125, USA\\
$^{3}$University of North Carolina at Chapel Hill, 120 E. Cameron Ave., Chapel Hill, NC 27514, USA\\
$^{4}$Sydney Institute for Astronomy, School of Physics, The University of Sydney, NSW 2006, Australia\\
$^{5}$ARC Centre of Excellence for Gravitational Wave Discovery (OzGrav), Hawthorn, Victoria, 3122, Australia\\
$^{6}$Center for Gravitation, Cosmology, and Astrophysics, Department of Physics \& Astronomy, University of Wisconsin-Milwaukee, PO Box 413, Milwaukee, WI 53201, USA\\
$^{7}$Center for Astrophysics | Harvard \& Smithsonian, 60 Garden St, Cambridge, MA 02138, USA\\
$^{8}$Institute of Science and Technology Austria, Am Campus 1, 3400, Klosterneuburg, Austria\\
$^{9}$Department of Physics, University of Warwick, Gibbet Hill Road, Coventry CV4 7AL, UK 
}

\date{Accepted XXX. Received YYY; in original form ZZZ}

\pubyear{2026}

\begin{document}
\label{firstpage}
\pagerange{\pageref{firstpage}--\pageref{lastpage}}
\maketitle

\begin{abstract}
Wide-field time-domain radio surveys have recently unveiled a new class of Galactic radio emitter: \acp{lpt}.
Despite the rapid pace of discovery, their origins remain poorly understood, with two main theories emerging: slowly spinning magnetars or binary \acp{wd} with magnetic interactions.
All three \acp{lpt} with well-characterised optical counterparts are \acp{wd} in binary systems, but even within that class there is diversity, with one accreting and two detached systems.
In this Letter, we show that the \ac{lpt} \jf is a \ac{cv}---a \ac{wd} accreting from a low-mass companion---with a spectroscopic orbital period consistent with the radio pulse period, and specifically, potentially consistent with being a polar.
Unlike the other known \ac{lpt} \ac{cv}, \jf does not exhibit the \ion{He}{II} emission that is a hallmark of magnetic \acp{cv} in their high accretion states.
We also present new optical time-series and infrared photometric observations, showing flickering at the 10\% level and suggesting a possible period bouncer origin.
This identification of \jf adds to the small sample of characterised \ac{lpt} counterparts, and shows that \ac{lpt} emission can persist in an accreting \ac{wd} binary at a low accretion rate.
\end{abstract}




\acresetall
\section{Introduction}

\begin{figure*}
	\includegraphics[width=\textwidth]{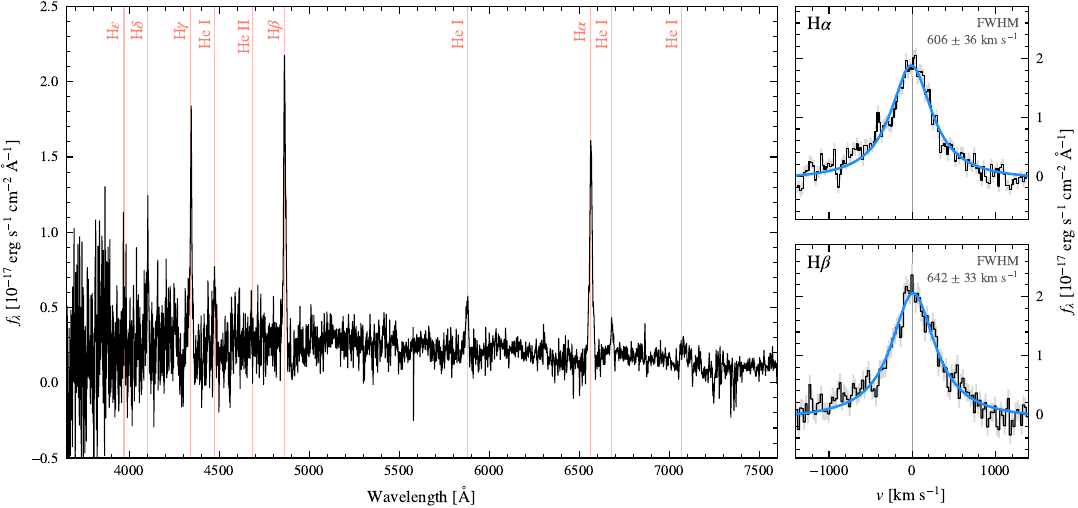}
    \caption{\textit{Left:} Coadded and smoothed spectrum of \jf from Magellan/MagE spectroscopy, with poorly calibrated data redward of $\sim7500$\,\AA\ truncated. The locations of prominent Balmer, \ion{He}{I}, and \ion{He}{II} lines are marked. Like many non-magnetic \acp{cv}, H$\alpha$, H$\beta$, and H$\gamma$ Balmer lines and \ion{He}{I} 4471\,\AA, 5876\,\AA, and 6678\,\AA\ lines are seen in emission. 
    We note the lack of \ion{He}{II} 4686\,\AA\ emission that is generally seen in high accretion state magnetic \acp{cv}.
    \textit{Right:} Rest-frame coadded line profiles of the H$\alpha$ and H$\beta$ emission lines, with Lorentzian fits in blue.
    }
    \label{fig:stacked_spectrum}
\end{figure*}

\Acp{lpt} are a new mysterious class of time-domain radio phenomenon, characterised by their bright, highly polarised, pulsed periodic radio emission.
They have periods spanning minutes to hours, with pulse durations from a few seconds to over 30\,min. 
With periods too slow for typical pulsars, yet radio luminosities much larger than flaring stars, ultracool dwarfs, and typical \acp{cv}, their discovery has challenged our understanding of Galactic radio emitters (e.g., \citealt{2005Natur.434...50H, 2022Natur.601..526H, 2024NatAs...8.1159C}; see \citealt{ReaLPT2026} for a review).

Two main classes of theories have emerged to explain \acp{lpt}: slowly rotating magnetars  \citep[e.g.,][]{2020MNRAS.496.3390B, 2022ApJ...934..184R, 2024MNRAS.533.2133C, 2026ApJ...996..141C} and magnetically interacting \ac{wd} binaries \citep[e.g.,][]{2025ApJ...981...34Q, 2026ApJ...999L...2Z, 2026ApJ...997..124Y, 2026arXiv260808243Q}.
Thus far, at least for the longer-period \acp{lpt} ($\gtrsim 1.3$\,hr), the \ac{wd} binary scenario has seen more observational support. 
As of late 2026, there are approximately 17 published \acp{lpt}.
Three have spectroscopically confirmed optical counterparts, all \acp{wd} in binaries with cool companions: GLEAM-X J0704$-$37 \citep{2024ApJ...976L..21H, 2025A&A...695L...8R} and ILT J1101$+$5521 \citep{2025NatAs...9..672D, 2026OJAp....967716R} are detached \ac{wd}--M dwarf binaries, while ASKAP J1745$-$5051 \citep{2026NatAs..10.1166R, 2026arXiv260628993K} is a magnetic \ac{wd} accreting from a low-mass donor (i.e., a magnetic \ac{cv}).

ASKAP J144834$-$685644 (hereafter \jf) was discovered by \citet{anumarlapudi_j1448} as an \ac{lpt} exhibiting emission across the electromagnetic spectrum, from radio through X-rays; we briefly describe their findings here.
While \jf was first identified in a search for circularly polarised radio sources, it has exhibited both linearly and circularly polarised radio emission, with a period of $\approx1.5$\,hr and pulse widths of $\approx30$\,min.
\jf is an X-ray source, albeit at much lower luminosities than \acp{cv}, magnetic or otherwise, at the same radio brightness (by at least two orders of magnitude).
Its UV--optical--infrared \ac{sed} peaks in the near-UV, and is reasonably well fit by a white dwarf with a cool companion.
Archival optical imaging shows that years before its discovery as an \ac{lpt}, \jf underwent an outburst, peaking $\approx 4$\,mag brighter than its quiescent level.
Subsequent simultaneous radio and optical observations have not seen another occurrence of such an outburst, demonstrating that the \ac{lpt} emission is not necessarily tied to an optically bright high state.
The distance to \jf is poorly constrained due to loose limits on the radio dispersion-measure (DM $< 720$\,pc\,cm$^{-3}$).
Based on the multiwavelength data, \citet{anumarlapudi_j1448} found it likely to be either an edge-on disk-dominated magnetic \ac{wd} binary or a detached \ac{wd} binary, although they could not rule out a transitional millisecond pulsar, an isolated magnetar, or an isolated \ac{wd} pulsar.

In this Letter, we present Magellan and Gemini optical and infrared observations of \jf, showing that it is a \ac{cv} with an orbital period consistent with the \ac{lpt} radio period.
This represents the second \ac{lpt}, after ASKAP J1745$-$5051 \citep{2026NatAs..10.1166R}, that has been confirmed as an accreting \ac{wd} binary, though we find \jf to lack the characteristic \ion{He}{II} emission that is associated with high accretion state magnetic \acp{cv}.
We describe the observations in Sec.~\ref{sec:obs}.
In Sec.~\ref{sec:components}, we model the physical characteristics of the system, including assessing whether its donor is an M dwarf or substellar.
We place these results in context in Sec.~\ref{sec:discussion}.
Unless otherwise specified, all magnitudes are in the AB system.

\section{Observations}
\label{sec:obs}

\subsection{Optical spectroscopy}
We obtained three hours of medium-resolution ($\mathrm{R} \approx 5800$) optical spectroscopy using the MagE spectrograph \citep{mage_instrument} on the 6.5\,m Magellan Baade telescope at Las Campanas Observatory on UT 2026-07-09.
Using the 0.7\arcsec\ slit, we took a continuous series of 600\,s exposures using the fast readout mode which has a 21\,s readout time.
We reduced the data with \texttt{PypeIt} \citep{Prochaska20_pypeit_python} using standard routines.
Due to intermittent clouds, only the latter two hours of observations were usable.
Of those, the underlying continuum trace is barely detected, while emission lines are clearly visible in each exposure.

\subsection{Optical photometry}
We imaged \jf\  in the $g$ and $r$ bands for approximately two hours each using the GMOS imaging spectrograph \citep{2004PASP..116..425H, 2016SPIE.9908E..2SG} on the 8\,m Gemini South telescope on UT 2026-03-13 (Program GS-2026A-Q-208, PI: Anumarlapudi).
Exposures lasted 120\,s in the $r$ band and 90\,s in the $g$ band, with a 29\,s readout between exposures. 
The data were reduced using the \textsc{dragons} pipeline \citep{dragons} to perform bias subtraction, flat fielding, and astrometric calibration. 
We used the \texttt{photutils} \citep{2026zndo..19636730B} implementation of DAOFind \citep{1987PASP...99..191S} to perform source extraction.
Data from the Dark Energy Camera Plane survey \citep[DECaPs;][]{decaps} were used for photometric calibration.
\jf is clearly detected in the individual exposures with \ac{snr} $\approx 39$ in $g$ and $32$ in $r$.

\subsection{Infrared photometry}
We observed \jf\ in the $K_S$ band using the FourStar near-infrared imager on the 6.5\,m Magellan Baade telescope on UT 2026-02-01.
We used a sequence of dithered exposures totalling 21.8\,min of science integration, and reduced the data using FourCLift \citep{2014ApJ...783..110K}.
No source is detected at the location of \jf in the stacked image, with a 5$\sigma$ limiting magnitude of $K_S > 22.9$\,mag.

\section{The components of the \jf\ binary}
\label{sec:components}

\subsection{Spectra and radial velocity variations}

We show the coadded and smoothed optical spectrum of \jf in Figure~\ref{fig:stacked_spectrum}.
H$\alpha$, H$\beta$, and H$\gamma$ Balmer emission features are clearly present, along with weaker \ion{He}{I} emission at 4471\,\AA, 5876\,\AA, and 6678\,\AA. 
Notably, there is a lack of \ion{He}{II} 4686\,\AA\ emission (with \ion{He}{II}/H$\beta < 0.1$), in contrast to magnetic \acp{cv} in high accretion states, where photoionisation from the shock-heated accretion column results in prominent \ion{He}{II} emission \citep{1995cvs..book.....W}.
In comparison, ASKAP J1745$-$5051, the other known \ac{lpt} \ac{cv}, displays clear \ion{He}{II} emission, with \ion{He}{II}/H$\beta \approx 0.4$ \citep{2026NatAs..10.1166R}.
We see no evidence for donor absorption lines (e.g., \ion{Ca}{II}, \ion{K}{I}, or \ion{Na}{I}), unsurprising given the very weakly detected continuum.

\begin{figure}
	\includegraphics[width=\columnwidth]{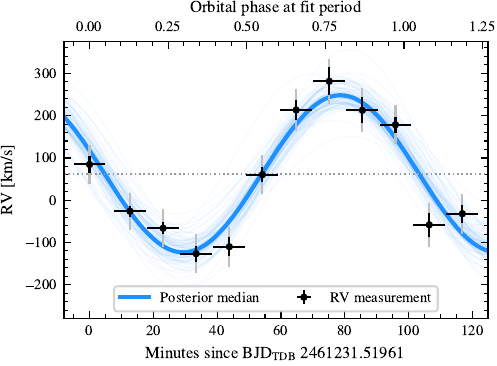}
    \caption{Radial velocities of \jf from Magellan/MagE spectroscopy, measured from joint fitting of the H$\alpha$ and H$\beta$ emission lines. Exposure durations are represented by the horizontal black error bars, formal \ac{rv} uncertainties by the vertical black error bars, and those uncertainties with the fitted jitter added in quadrature by the grey error bars. The dotted grey line shows the barycentric systemic velocity. The posterior median \ac{rv} fit is shown in the thick blue line, with a sample of other posterior draws in light blue. Due to the large uncertainties in the radio ephemeris and period, we are unable to determine the orbital phase at which the radio pulses are emitted.}
    \label{fig:rv}
\end{figure}

To measure the \ac{rv} variations, we simultaneously fit the
H$\alpha$ and H$\beta$ emission lines from each 600\,s exposure with two Lorentzian profiles that share an \ac{rv}.\footnote{Fitting the \ion{He}{I} lines gives similar results.}
We then fit a circular Keplerian orbit to the \acp{rv} using the \texttt{emcee} Markov chain Monte Carlo sampler \citep{2013PASP..125..306F}.
The five free parameters are the orbital period $P$, \ac{rv} semi-amplitude $K$, barycentric systemic velocity $\gamma$, logarithm of the \ac{rv} jitter $\ln{s}$, and epoch of maximum \ac{rv}.
We adopted uniform priors on all five parameters, and used 32 walkers taking 55,000 steps, discarding the first 5000 as burn-in.

\begin{table}
\centering
\caption{Circular Keplerian fit to the H$\alpha$ and H$\beta$ radial velocities of \jf.
Values are posterior medians with 68\% credible intervals. Reduced $\chi^2$ = 1.18 including jitter.}
\label{tab:rv_fit}
\renewcommand{\arraystretch}{1.4}
\begin{tabular}{llc}
\toprule
Parameter & Unit & Value \\
\midrule
\midrule
Orbital period $P$        & min             & $98.6^{+4.2}_{-4.1}$ \\
RV semi-amplitude $K$     & km/s     & $186^{+21}_{-22}$ \\
Barycentric systemic velocity $\gamma$& km/s     & $63^{+14}_{-15}$ \\
RV jitter $s$             & km/s     & $42^{+17}_{-12}$ \\
Epoch of maximum RV & BJD$_{\rm TDB}$ & $2461231.5742 \pm 0.0013$ \\

\bottomrule
\end{tabular}
\end{table}

The \ac{rv} measurements and fit are shown in Figure~\ref{fig:rv}, with the posteriors described in Table~\ref{tab:rv_fit}.
We find a \ac{rv} semi-amplitude $K = 186^{+21}_{-22}$\,km/s, with an orbital period $P = 98.6^{+4.2}_{-4.1}$\,min, confirming the binary nature of \jf.
The semi-amplitude $K$ does not appear to be tracing either the \ac{wd} or the donor (see Sec.~\ref{sec:emission-origin}).
The spectroscopic period is consistent with the radio-derived period of 93.8513(8)\,min to within 1.2$\sigma$ (see Appendix \ref{app:timing} for caveats on the radio period).
Assuming that the radio period is equal to the orbital period, this suggests that the radio pulses are modulated by the orbit, though emission at a beat period between the orbital period and (unknown) \ac{wd} spin period cannot be ruled out. 
Since the optical and the radio data are separated by $>$2\,years, the lack of a phase-connected timing solution means that we cannot precisely constrain the orbital phase at which the radio pulses are emitted. 
In addition, even with the current timing solution, the uncertainty on the predicted phase for radio pulses is very large ($\approx$11\% of the orbit; see Appendix \ref{app:timing}) given the wide pulses, and hence deriving meaningful constraints on the orbital phase of the radio pulses is difficult.

\begin{figure}
	\includegraphics[width=\columnwidth]{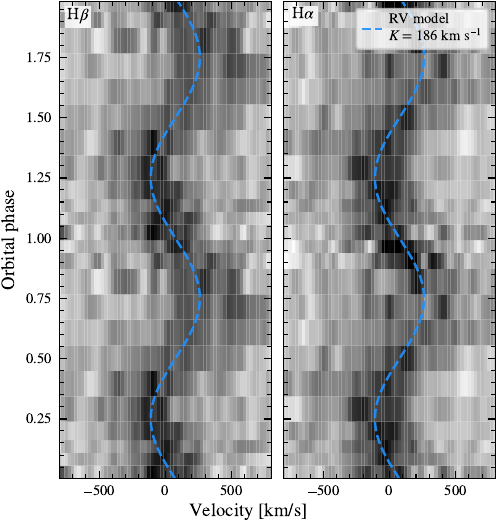}
    \caption{Trailed spectrogram of the H$\beta$ and H$\alpha$ lines from the individual spectra, phase-folded to the 98.6\,min fit period. Two cycles are shown for ease of viewing.
    The posterior median \ac{rv} model is overlaid in the dashed blue line.}
    \label{fig:trailed_spectrogram}
\end{figure}

\subsection{The origin of the emission lines}
\label{sec:emission-origin}

At this orbital period, the \ac{rv} semi-amplitude due to the reflex motion of the \ac{wd} from Kepler's third law scales as 
\begin{equation}
    K = \mathrm{58.9\,km/s} \times \sin{i} \Big(\frac{M_2}{0.106\,\mathrm{M}_\odot}\Big) \Big(\frac{M_1 + M_2}{0.8\,\mathrm{M}_\odot +0.106\,\mathrm{M}_\odot}\Big)^{-2/3},
\end{equation}
for the inclination \textit{i}, \ac{wd} mass $M_1$, and donor mass $M_2$.
Even at edge-on viewing angles, the unequal mass ratio means that any emission whose centroid follows the \ac{wd}'s orbit, such as the \ac{wd} photosphere or a disk, should have $K \lesssim 60$\,km/s, compared to the observed $K = 186^{+21}_{-22}$\,km/s.
Thus, the Balmer and \ion{He}{I} emission cannot arise from the \ac{wd} or from a symmetric disk around it.

As noted above, we fit the Balmer emission lines with Lorentzian line profiles.
In polars (magnetic \acp{cv} with magnetic fields strong enough to prevent the formation of a disk), the emission lines often consist of a broad base component from the magnetically channelled accretion flow and a narrow component from the donor \citep{1977ApJ...212L.121C}.
However, using multiple-component models did not substantially improve the fits to our data, with poor agreement in the fit widths between H$\alpha$ and H$\beta$ and no evidence of a narrow ($<100$\,km/s) component in either.
The single-component nature of the Balmer emission is also illustrated in the H$\alpha$ trailed spectrogram (Figure~\ref{fig:trailed_spectrogram}).
The line is broad and follows a single sinusoidal track with no apparent second component. 
This is also evidence against the emission coming from a hotspot on the disk: with no underlying slower broad component, there is unlikely to be a disk at all over which a hotspot forms.
For comparison, in the well-studied 82\,min \ac{cv} WZ Sge, the hotspot contributes 50\% of the line flux at maximum \citep{2000MNRAS.318..440M}.

Both H$\alpha$ and H$\beta$ have $\mathrm{FWHM} \approx 620$\,km/s, measured from Lorentzian fits to an \ac{rv}-corrected  (rest frame) coadded spectrum of all the epochs (see Figure~\ref{fig:stacked_spectrum}).
The lines are too broad to be from the irradiated face of the donor, which typically has emission lines with $\mathrm{FWHM} \approx 100$\,km/s \citep{1985ASSL..113..151L}.

Using the line ratios, we can estimate the optical depth and gas density of the line-emitting region.
From an \ac{rv}-corrected coadded spectrum with an extinction correction ($A_V = 0.3\pm 0.3$ to account for the unknown distance), we measure $\mathrm{H}\alpha / \mathrm{H}\beta = 1.07 \pm 0.14$ and
$\mathrm{H}\gamma / \mathrm{H}\beta = 0.61 \pm 0.05$.
The ratios are largely typical for \acp{cv}, with $\mathrm{H}\gamma / \mathrm{H}\beta$ slightly on the lower side \citep{1983ApJS...53..523W, 1988MNRAS.233..513E, 2026A&A...712A.116H}.
These values suggest that the emitting gas has high optical depths with temperatures of roughly $10^4$\,K, and hydrogen number density $N_0 \approx 10^{12}$--$10^{13}$\,cm$^{-3}$, typical for  \ac{cv} accretion structures \citep{1980ApJS...42..351D, 1991AJ....101.1929W}.
At these number densities, the plasma frequency is $\approx9$--28\,GHz for fully ionised gas, far above the 799--1090\,MHz band in which ASKAP discovered \jf.
The coherent radio emission must therefore arise in a lower-density region and propagate along a sight-line that avoids the line-emitting gas.

Since we have ruled out the \ac{wd} photosphere, a symmetric disk around the \ac{wd}, a hotspot on the disk, and the irradiated donor face, we conclude that the line emission is most likely to originate from an extended accretion stream between the donor and the \ac{wd}.

\subsection{Constraints on the donor and distance}

\begin{figure}
	\includegraphics[width=\columnwidth]{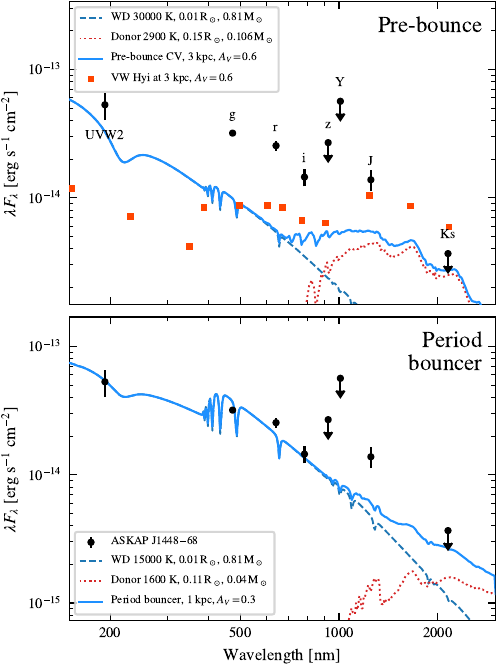}
    \caption{\ac{uv} through near-infrared \ac{sed} of \jf, with detections and 5$\sigma$ upper limits (arrows) shown in black. \textit{Top:} a pre-bounce \ac{wd}--M dwarf model (light blue) with a hot 30,000\,K \ac{wd} and a 2900\,K M dwarf donor, at 3\,kpc and $A_V=0.6$. In orange squares is the \ac{sed} of the pre-bounce 107\,min \ac{cv} VW Hyi, scaled to the same distance and extinction. 
    \textit{Bottom:} a period bouncer model (light blue), with a 15,000\,K \ac{wd} and a 1600\,K substellar donor, at a distance of 1\,kpc and $A_V = 0.3$, showing much better agreement with the data.}
    \label{fig:sed}
\end{figure}

\acp{cv} have well-understood evolutionary tracks.
Young \acp{cv} lose angular momentum via magnetic braking and general relativistic decay, shrinking their orbits until they reach the \ac{cv} period minimum at $P_{\rm orb} \approx 80$\,min \citep{2009MNRAS.397.2170G}.
At this point, the donor starts to become degenerate and expands in radius as it continues to lose mass, causing the \ac{cv} to evolve back towards longer periods as a ``period bouncer'' \citep{1995cvs..book.....W, 2011ApJS..194...28K}.
Thus \ac{cv} periods do not uniquely determine their evolutionary state; a \ac{cv} at a given period could either be evolving towards the period minimum or ``bouncing'' out.

Since \jf is a \ac{cv}, we consult the semi-empirical \ac{cv} donor sequence \citep[``optimal'' results from][]{2011ApJS..194...28K}. 
Assuming the system is pre-bounce (i.e., inspiralling), for a 1.64\,hr orbital period, the donor mass $M_2 \approx 0.106$\,M$_\odot$ and donor radius $R_2 \approx 0.15$\,R$_\odot$, with $T_{\rm eff} \approx 2900$\,K and absolute Vega magnitude $M_K = 8.8$.
There is no robust distance estimate from the radio observations, but given our 5$\sigma$ limiting magnitude of $K_S > 22.9$\,mag (AB) = 21.0\,mag (Vega), we find that \jf must be at a distance of $\gtrsim 2.7$\,kpc if it has an M dwarf donor.

If, however, \jf is a period bouncer, the donor mass could be significantly smaller. 
The post-bounce models presented in \citet{2011ApJS..194...28K} do not extend to periods as long as 1.64\,hr, but comparing to the known period bouncer SRGeJ0411 \citep{2024MNRAS.528..676G} with a very similar period, the donor mass could be as small as $M_2 \lesssim 0.04$\,M$_\odot$, with $R_2 \lesssim 0.11$\,R$_\odot$ and $T_\mathrm{eff} \lesssim 1800$\,K.
In that case, its bolometric luminosity $L_{\rm bol} \leq 1.1\times10^{-4}$\,L$_\odot$, corresponding to an absolute magnitude $M_{\rm bol} \geq 14.7$.
Taking a $K$-band bolometric correction of BC$_{\rm K} = 3.3$ from \citet{2004AJ....127.3516G}, this results in $M_K \geq 11.4$, or a distance constraint of $\gtrsim 830$\,pc.\footnote{A fainter donor could be closer.}

We plot the \ac{sed} of \jf from \citet{anumarlapudi_j1448} in Figure~\ref{fig:sed}, updated with our $K_S$ upper limit.
From line fits of the spectrum, we find that the emission line contribution to the broadband flux is $\lesssim 0.15$\,mag, so we neglect it for this analysis.
In the top panel of Figure~\ref{fig:sed}, we show a model of a pre-bounce \ac{cv}, composed of a BT-DUSTY \citep{2011ASPC..448...91A} model M dwarf donor and a 30,000\,K DA \ac{wd} atmosphere computed with OpenWD\footnote{\url{https://github.com/kareemelbadry/OpenWD}} (El-Badry et al., in prep). 
The system is placed at 3\,kpc to match the donor sequence distance constraint. 
Here, 3D extinction models indicate $A_V=0.6$ \citep{2025ApJ...992...39Z}.
By enforcing that the donor obeys the $K_S$ limit, we find that the \ac{wd} accretor is too faint for the \ac{uv} and optical \jf photometry, even if the \ac{wd} temperature is as high as 30,000\,K (much hotter than typical \ac{cv} accretors; \citealt{2022MNRAS.510.6110P}).
For comparison, we place the well-studied 107\,min pre-bounce \ac{cv} VW Hyi (\ac{wd} $T_{\rm eff} \approx$ 20,000\,K; \citealt{1996ApJ...471L..41S, 2006MNRAS.369.1537S}) at the same distance and extinction.
This, too, does not match the \jf data: it is too faint in the optical and \ac{uv}, and too bright in the $K_S$ near-infrared band.

We qualitatively find better results in the period bouncer case: using a cooler 15,000\,K \ac{wd} and a representative substellar donor ($T_{\rm eff} = 1600$\,K, 0.11\,R$_\odot$, 0.04\,M$_\odot$, and solar metallicities), we assume $A_V = 0.3$ and place the system at 1\,kpc.
This model largely reproduces the observed fluxes.
The $J$-band observation at 1250\,nm shows the largest deviation from the model; this could be due to cyclotron humps from strong magnetic fields \citep[e.g.,][]{2006AJ....131.2216H, 2008ApJ...672..531C}.
Lacking an independent measurement of distance, we are unable to confidently constrain whether or not \jf is a period bouncer from the photometry alone.
Spectroscopic observations with the \textit{James Webb Space Telescope} could definitively distinguish between an M dwarf or substellar donor and thus also confirm a distance.

\subsection{Photometric variability and inclination}

\begin{figure}
	\includegraphics[width=\columnwidth]{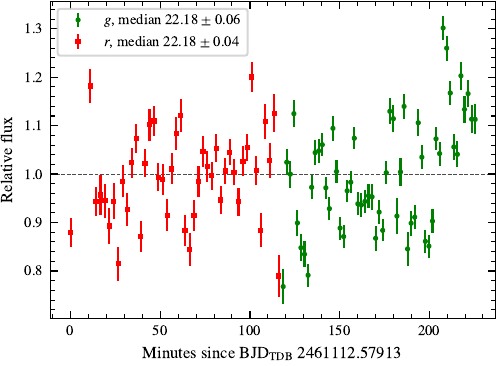}
    \caption{Light curve from Gemini/GMOS observations in SDSS $g$ and $r$ bands. Flickering at the $\approx10\%$ level is clearly seen.}
    \label{fig:lc}
\end{figure}

We show the Gemini/GMOS light curve of \jf in Figure~\ref{fig:lc}.
To calibrate out atmospheric transparency effects, we performed differential photometry relative to an ensemble of $\approx90$ comparison stars.
The median magnitudes over each sequence are $g = 22.18 \pm 0.06$ and $r = 22.18 \pm 0.04$, consistent with the quiescent DECam magnitudes ($g = 22.15 \pm 0.04$, $r = 22.06 \pm 0.09$;  \citealt{anumarlapudi_j1448}).
Both bands clearly show variability, with standard deviations of $\sigma_g$ = 12\% and $\sigma_r$ = 9\%.
Given the limited baseline ($\approx$ 1.2 orbital periods per band), the variability does not appear to be periodic. 
We performed Lomb-Scargle periodicity searches \citep{1976Ap&SS..39..447L, 1982ApJ...263..835S} over the $g$, $r$, and combined $g+r$ photometry. 
No peak exceeded the 10\% false-alarm level \citep{2008MNRAS.385.1279B} in any search; we thus do not detect any significant periodicity.
Apparently stochastic ``flickering'' is a well-known but somewhat poorly understood phenomenon associated with accretion, and has been thoroughly characterised in \acp{cv} \citep[e.g.,][]{2021MNRAS.503..953B, 2022MNRAS.516.5209I}.
This, in addition to the Balmer emission, further supports the presence of accretion in \jf.

In addition, there are no clear eclipses seen in the light curve.
Assuming a \ac{wd} mass of 0.8\,M$_\odot$ (typical for \acp{cv}; \citealt{2022MNRAS.510.6110P}) and using the period bouncer scenario donor mass of 0.04\,M$_\odot$, Kepler's third law gives an orbital separation $a = 0.66$\,R$_\odot$.
Assuming the $g$ and $r$ optical emission is dominated by the \ac{wd} (and not e.g., the accretion stream), the absence of eclipses requires $R_2 < a\cos{i}$, resulting in a maximum inclination $i_{\rm max} \leq 80 ^\circ$.\footnote{This constraint is very weakly dependent on the possible \ac{wd} and donor parameters; for combinations of $M_1 = 0.6-0.8$\,M$_\odot$, $M_2 = 0.04-0.106$\,M$_\odot$, and $R_2 = 0.1-0.15$\,R$_\odot$, $i_{\rm max}$ only varies from $76^\circ \leq i_{\rm max} \leq 82^\circ$.}
If the emission is instead largely from the accretion stream, the constraint strengthens, making $i \leq 80 ^\circ$ a conservative upper limit.
This is in slight conflict with the edge-on accretion disk interpretation from \ac{sed} fitting proposed by \citet{anumarlapudi_j1448}, suggesting that there indeed may not be a disk.

\section{Discussion and conclusions}
\label{sec:discussion}

\subsection{What kind of \ac{cv} is \jf?}
The Balmer and \ion{He}{I} emission lines, along with the $98.6^{+4.2}_{-4.1}$\,min orbital period, point strongly at a \ac{cv} nature for \jf.
However, the typical observables used to determine the magnetic nature of a \ac{cv} are absent in our data.
Chief among these is the lack of detectable \ion{He}{II} emission, indicating the absence of the high-ionisation region produced by magnetically-channelled accretion onto the \ac{wd} magnetic poles that is generally associated with magnetic \acp{cv}.
The very weak continuum in our observed spectra means that neither cyclotron humps nor any Zeeman splitting of the \ac{wd} Balmer absorption lines can be measured. 
So while we cannot place a limit on the \ac{wd} magnetic field, we can determine that \jf is not a high-accretion-state polar or \ac{ip}.

However, the detection of coherent, highly polarised radio emission points to an emission mechanism operating in a strongly magnetised plasma and therefore suggests the presence of a large-scale ordered magnetic field \citep[e.g.,][]{2017RvMPP...1....5M, 2021ApJ...922..166L}.
We consider magnetic \ac{cv} states compatible with our observations.
One clue lies in the archival DECam observations presented in \citet{anumarlapudi_j1448}.
These show that a few years prior to its radio discovery as an \ac{lpt}, \jf exhibited a flaring or high state lasting at least two months, with optical detections 4 mag brighter than its quiescent state. 
The high state has not been seen since \jf's discovery as an \ac{lpt}, even during follow-up optical observations by \citet{anumarlapudi_j1448} that are near-contemporaneous with the radio observations.
This suggests that the radio \ac{lpt} emission is not necessarily tied to an optically bright high state.
Polars such as EF Eri have been observed to display multiple states with similar magnitude differences and timescales, showing spectroscopic features ranging from narrow chromospheric Balmer emission and no \ion{He}{II} in the low state to broad Balmer emission and strong \ion{He}{II} in the high state \citep{1998ASPC..137..446W, 2000A&A...354L..49B, 2006ApJ...652..709H, 2025MNRAS.544..309K}.
In \jf, the observational evidence, primarily the lack of a bright accretion disk and \ion{He}{II} emission, yet emission lines too broad to be from just the irradiated donor, suggests that it may be a polar between the typical ``low'' and ``high'' accretion states.
Long-term spectroscopic monitoring could measure and characterise any changes in its accretion state.

Could \jf be a period bouncer, as has been proposed for the \ac{lpt} ASKAP J1745$-$5051 \citep{2026NatAs..10.1166R, 2026arXiv260628993K}?
Without a distance measurement, it is difficult to use our existing data to confidently measure the donor's mass, though our \ac{sed} fitting (Figure~\ref{fig:sed}) suggests that the \jf photometry is more compatible with a period bouncer explanation than otherwise.
Deeper infrared observations could better characterise the donor's emission, allowing for a more robust determination of its nature and the system's distance.

In comparing to several benchmark \acp{cv}, \citet{anumarlapudi_j1448} found that \jf's X-ray to \textit{J}-band flux ratio was suppressed by a few orders of magnitude.\footnote{We note that such analyses will be influenced by the presence or absence of \textit{J}-band cyclotron humps that depend on the magnetic field strength, which is currently unmeasured for \jf.}
With \jf's \ac{cv} determination now secure, we reevaluate this claim.
\jf has an X-ray to \textit{J} flux ratio $(\nu F_\nu)_X / (\nu F_\nu)_J \approx$~1.1-3.3 \citep[using unabsorbed XMM-Newton fluxes from][]{anumarlapudi_j1448}.
While the prototypical polar AM Her has $(\nu F_\nu)_X / (\nu F_\nu)_J = 7.7$ in its high state \citep{1982ApJ...257..686S, 2020A&A...642A.134S}, it falls to $(\nu F_\nu)_X / (\nu F_\nu)_J = 0.18$ in its low state \citep{2008ApJ...683..409C, 2020A&A...642A.134S}.
The polar period bouncers WD J1820 and WD J1907 have $(\nu F_\nu)_X / (\nu F_\nu)_J = 0.03$ and 0.26 respectively \citep{2025MNRAS.540..633C}.
U Gem, an example of a non-magnetic \ac{cv}, shows $(\nu F_\nu)_X / (\nu F_\nu)_J = 0.14$ in quiescence \citep{1983MNRAS.204.1105B, 2006MNRAS.372..450G}.
Using the unabsorbed fluxes from \citet{anumarlapudi_j1448}, a fiducial distance of $d = 1$\,kpc results in an isotropic X-ray luminosity of $L_X \approx 3 \times 10^{30} (d /\mathrm{1\,kpc})^2$\,erg/s, lying toward the lower end of the \ac{cv} $L_X$ distribution \citep{2025PASP..137a4201R}.
In summary, \jf appears to have an X-ray to \textit{J}-band flux ratio larger than most low-state \acp{cv}, but not reaching that of high-state polars, again indicating that it might be undergoing a moderate level of accretion.

\subsection{\jf in its \ac{lpt} context}

Of its \ac{lpt} siblings, \jf is most similar to ASKAP J1745$-$5051, a magnetic \ac{cv} that could be a period bouncer \citep{2026NatAs..10.1166R, 2026arXiv260628993K}.
ASKAP J1745$-$5051 has a spectroscopically confirmed 1.37\,hr period, similar to but shorter than \jf's 1.64\,hr, and both are detected in X-rays \citep{2026NatAs..10.1166R, 2026A&A...710L..27I}.
On the other hand, ASKAP J1745$-$5051 shows \ion{He}{II} emission and displays Balmer features which are almost twice as broad as those from \jf.
This suggests that the line-emitting regions in the two systems are different; \jf's lower ionisation and narrower FWHM may point to emission originating farther from the \ac{wd}'s magnetic poles, painting an overall picture of a lower accretion rate.

One of the most valuable observables for ascertaining the origin of the \ac{lpt} emission is the orbital phase at which the radio pulses are emitted.
Of the other \ac{wd} binary \acp{lpt}, the detached \ac{wd}--M dwarf systems ILT J1101 \citep{2025NatAs...9..672D} and GLEAM-X J0704$-$37 \citep{2024ApJ...976L..21H} emit their radio pulses just after maximum M dwarf \ac{rv} \citep{2026OJAp....967716R}, whereas the magnetic \ac{cv} ASKAP J1745$-$5051 \citep{2026NatAs..10.1166R} emits most of its pulses after conjunction (when the two components are along our line-of-sight).
Unfortunately, due to \jf's irregular and extremely long radio pulses (lasting a third of the orbital period), the radio period and ephemeris are poorly measured, preventing meaningful phase-folding with the optical observations (see Appendix~\ref{app:timing}).
Future simultaneous radio and optical observations could avoid this complication.

While there is no firm optical counterpart, the \ac{lpt} GPM J1839$-$10 \citep{2023Natur.619..487H} has been recently proposed by \citet{2026NatAs..10..522H} to also be a \ac{wd} with a low-mass stellar companion, based on decades of radio timing.
In their model, the radio pulses are produced at the 22\,min beat period between a slightly longer \ac{wd} spin period and an 8.75\,hr orbital period.
Though \jf has a much shorter orbital period, the \ac{wd} spin period is unknown, allowing for the possibility that the radio pulses are emitted at a spin--orbit beat period, like GPM J1839$-$10.
While polars generally have \ac{wd} spins synchronized to their orbits, a nova eruption could temporarily desynchronize the system, resulting in a transient asynchronous polar phase \citep[e.g.,][]{1988ApJ...332..282S, 2007A&A...473..511S}.
Further spectroscopic observations will improve our measurement of the orbital period, revealing any discrepancy between the radio and orbital periods to probe the possibility that \jf is an asynchronous polar.

All the confirmed \ac{lpt} optical counterparts are binary systems involving a \ac{wd} and a low-mass cool companion; nevertheless, this characterisation of \jf as a \ac{cv} adds to their diversity. 
There are two detached \ac{wd}--M dwarf systems \citep{2025A&A...695L...8R, 2026OJAp....967716R}, a magnetic \ac{cv} \citep{2026NatAs..10.1166R, 2026arXiv260628993K} likely in a high state, and now a \ac{cv} where any magnetic field, if present, is not driving high-state accretion.
As the number of discovered \acp{lpt} continues to grow, multiwavelength characterisation will be crucial to determining their physical properties and solving the mystery of their origins.

\section*{Acknowledgements}

GM is supported by the Brinson Foundation through the Brinson Prize Fellowship Program. IdR is supported by the Australian Research Council Centre of Excellence for Gravitational Wave Discovery (OzGrav), project number CE230100016. DLK is  supported by NSF grant AST-2511757.
This paper includes data gathered with the 6.5 meter Magellan Telescopes located at Las Campanas Observatory, Chile. 
This work made use of 
\texttt{numpy} \citep{harris2020array}, 
\texttt{astropy} \citep{astropy:2013}, 
\texttt{matplotlib} \citep{Hunter:2007}, 
\texttt{pandas} \citep{mckinney2010data}, 
\texttt{scipy} \citep{2020SciPy-NMeth}, 
\texttt{pypeit} \citep{Prochaska20_pypeit_python}, 
\texttt{lmfit} \citep{2021zndo....598352N}, 
\texttt{emcee} \citep{2013PASP..125..306F},
Vizier \citep{vizier},
and 
\texttt{OverCite} \citep{Shariat26_overcite}.
The large language model Anthropic Claude (Opus 5.5) was used in the preparation of this manuscript to assist with data processing and in writing code for analysis and plotting.
We also benefited from discussions with Claude about the results and interpretation.
The authors retain full responsibility for the contents of the manuscript.


\section*{Data Availability}

The spectroscopic and photometric data will be provided by the authors upon reasonable request.



\bibliographystyle{mnras}
\bibliography{lpt} 

@ARTICLE{ReaLPT2026,
       author = {{Rea}, Nanda and {Hurley-Walker}, Natasha and {Caleb}, Manisha},
        title = "{Long period transients (LPTs): A comprehensive review}",
      journal = {Journal of High Energy Astrophysics},
         year = 2026,
        month = apr,
       volume = {52},
          eid = {100566},
        pages = {100566},
          doi = {10.1016/j.jheap.2026.100566},
archivePrefix = {arXiv},
       eprint = {2601.10393},
 primaryClass = {astro-ph.HE},
       adsurl = {https://ui.adsabs.harvard.edu/abs/2026JHEAp..5200566R}
}

@ARTICLE{anumarlapudi_j1448,
       author = {{Anumarlapudi}, Akash and {Kaplan}, David L. and {Rea}, Nanda and {Erasmus}, Nicolas and {Kelson}, Daniel and {Ocker}, Stella Koch and {Lenc}, Emil and {Dobie}, Dougal and {Hurley-Walker}, Natasha and {Sivakoff}, Gregory and {Buckley}, David A.~H. and {Murphy}, Tara and {Pritchard}, Joshua and {Driessen}, Laura and {Rose}, Kovi and {Zic}, Andrew},
        title = "{ASKAP J144834{\ensuremath{-}}685644: a newly discovered long period radio transient detected from radio to X-rays}",
      journal = {\mnras},
         year = 2025,
        month = sep,
       volume = {542},
       number = {2},
        pages = {1208-1232},
          doi = {10.1093/mnras/staf1227},
archivePrefix = {arXiv},
       eprint = {2507.13453},
 primaryClass = {astro-ph.HE},
       adsurl = {https://ui.adsabs.harvard.edu/abs/2025MNRAS.542.1208A}
}

@INPROCEEDINGS{mage_instrument,
       author = {{Marshall}, J.~L. and {Burles}, Scott and {Thompson}, Ian B. and {Shectman}, Stephen A. and {Bigelow}, Bruce C. and {Burley}, Gregory and {Birk}, Christoph and {Estrada}, Jorge and {Jones}, Patricio and {Smith}, Matthew and {Kowal}, Vince and {Castillo}, Jerson and {Storts}, Robert and {Ortiz}, Greg},
        title = "{The MagE spectrograph}",
    booktitle = {Ground-based and Airborne Instrumentation for Astronomy II},
         year = 2008,
       editor = {{McLean}, Ian S. and {Casali}, Mark M.},
       series = {Society of Photo-Optical Instrumentation Engineers (SPIE) Conference Series},
       volume = {7014},
        month = jul,
          eid = {701454},
        pages = {701454},
          doi = {10.1117/12.789972},
archivePrefix = {arXiv},
       eprint = {0807.3774},
 primaryClass = {astro-ph},
       adsurl = {https://ui.adsabs.harvard.edu/abs/2008SPIE.7014E..54M}
}

@ARTICLE{2026NatAs..10.1166R,
       author = {{Rose}, Kovi and {Pritchard}, Joshua and {Murphy}, Tara and {Driessen}, L.~N. and {Kaplan}, D.~L. and {Caleb}, M. and {Wang}, Ziteng and {Zic}, A. and {Andreoni}, I. and {Carney}, J. and {Barlow}, B.~N. and {Dobie}, D. and {Gu}, M. and {Heald}, G. and {Huber}, D. and {Lenc}, E. and {Leung}, J.~K. and {Lu}, W. and {Momose}, R. and {Pedersen}, M.~G. and {Qu}, Y. and {Rea}, N. and {de Ruiter}, I. and {Shaji}, K. and {Sivakoff}, G.~R. and {Thomson}, A.~J.~M. and {Wang}, Y.~L. and {Yang}, G.~J. and {Zahedy}, F.},
        title = "{Periodic radio and X-ray emission from an accreting white dwarf binary}",
      journal = {Nature Astronomy},
         year = 2026,
        month = aug,
       volume = {10},
        pages = {1166-1178},
          doi = {10.1038/s41550-026-02882-x},
archivePrefix = {arXiv},
       eprint = {2606.04232},
 primaryClass = {astro-ph.HE},
       adsurl = {https://ui.adsabs.harvard.edu/abs/2026NatAs..10.1166R}
}

@ARTICLE{Prochaska20_pypeit_python,
       author = {{Prochaska}, J. and {Hennawi}, Joseph and {Westfall}, Kyle and {Cooke}, Ryan and {Wang}, Feige and {Hsyu}, Tiffany and {Davies}, Frederick and {Farina}, Emanuele and {Pelliccia}, Debora},
        title = "{PypeIt: The Python Spectroscopic Data Reduction Pipeline}",
      journal = {The Journal of Open Source Software},
         year = 2020,
        month = dec,
       volume = {5},
       number = {56},
          eid = {2308},
        pages = {2308},
          doi = {10.21105/joss.02308},
archivePrefix = {arXiv},
       eprint = {2005.06505},
 primaryClass = {astro-ph.IM},
       adsurl = {https://ui.adsabs.harvard.edu/abs/2020JOSS....5.2308P}
}

@ARTICLE{Shariat26_overcite,
       author = {{Shariat}, Cheyanne},
        title = "{OverCite: Add Citations in LaTeX without Leaving the Editor}",
      journal = {Research Notes of the American Astronomical Society},
         year = 2026,
        month = apr,
       volume = {10},
       number = {4},
          eid = {86},
        pages = {86},
          doi = {10.3847/2515-5172/ae5dbc},
archivePrefix = {arXiv},
       eprint = {2604.15366},
 primaryClass = {cs.DL},
       adsurl = {https://ui.adsabs.harvard.edu/abs/2026RNAAS..10...86S}
}

@ARTICLE{2004PASP..116..425H,
       author = {{Hook}, I.~M. and {J{\o}rgensen}, Inger and {Allington-Smith}, J.~R. and {Davies}, R.~L. and {Metcalfe}, N. and {Murowinski}, R.~G. and {Crampton}, D.},
        title = "{The Gemini-North Multi-Object Spectrograph: Performance in Imaging, Long-Slit, and Multi-Object Spectroscopic Modes}",
      journal = {\pasp},
         year = 2004,
        month = may,
       volume = {116},
       number = {819},
        pages = {425-440},
          doi = {10.1086/383624},
       adsurl = {https://ui.adsabs.harvard.edu/abs/2004PASP..116..425H}
}

@INPROCEEDINGS{2016SPIE.9908E..2SG,
       author = {{Gimeno}, German and {Roth}, Katherine and {Chiboucas}, Kristin and {Hibon}, Pascale and {Boucher}, Luc and {White}, John and {Rippa}, Matthew and {Labrie}, Kathleen and {Turner}, James and {Hanna}, Kevin and {Lazo}, Manuel and {P{\'e}rez}, Gabriel and {Rogers}, Rolando and {Rojas}, Roberto and {Placco}, Vinicius and {Murowinski}, Richard},
        title = "{On-sky commissioning of Hamamatsu CCDs in GMOS-S}",
    booktitle = {Ground-based and Airborne Instrumentation for Astronomy VI},
         year = 2016,
       editor = {{Evans}, Christopher J. and {Simard}, Luc and {Takami}, Hideki},
       series = {Society of Photo-Optical Instrumentation Engineers (SPIE) Conference Series},
       volume = {9908},
        month = aug,
          eid = {99082S},
        pages = {99082S},
          doi = {10.1117/12.2233883},
       adsurl = {https://ui.adsabs.harvard.edu/abs/2016SPIE.9908E..2SG}
}

@ARTICLE{2014ApJ...783..110K,
       author = {{Kelson}, Daniel D. and {Williams}, Rik J. and {Dressler}, Alan and {McCarthy}, Patrick J. and {Shectman}, Stephen A. and {Mulchaey}, John S. and {Villanueva}, Edward V. and {Crane}, Jeffrey D. and {Quadri}, Ryan F.},
        title = "{The Carnegie-Spitzer-IMACS Redshift Survey of Galaxy Evolution since z = 1.5. I. Description and Methodology}",
      journal = {\apj},
         year = 2014,
        month = mar,
       volume = {783},
       number = {2},
          eid = {110},
        pages = {110},
          doi = {10.1088/0004-637X/783/2/110},
archivePrefix = {arXiv},
       eprint = {1402.1771},
 primaryClass = {astro-ph.CO},
       adsurl = {https://ui.adsabs.harvard.edu/abs/2014ApJ...783..110K}
}

@BOOK{1995cvs..book.....W,
       author = {{Warner}, Brian},
        title = "{Cataclysmic variable stars}",
         year = 1995,
       volume = {28},
       series = {},
       adsurl = {https://ui.adsabs.harvard.edu/abs/1995cvs..book.....W}
}

@ARTICLE{2024ApJ...976L..21H,
       author = {{Hurley-Walker}, N. and {McSweeney}, S.~J. and {Bahramian}, A. and {Rea}, N. and {Horv{\'a}th}, C. and {Buchner}, S. and {Williams}, A. and {Meyers}, B.~W. and {Strader}, Jay and {Aydi}, Elias and {Urquhart}, Ryan and {Chomiuk}, Laura and {Galvin}, T.~J. and {Coti Zelati}, F. and {Bailes}, Matthew},
        title = "{A 2.9 hr Periodic Radio Transient with an Optical Counterpart}",
      journal = {\apjl},
         year = 2024,
        month = dec,
       volume = {976},
       number = {2},
          eid = {L21},
        pages = {L21},
          doi = {10.3847/2041-8213/ad890e},
archivePrefix = {arXiv},
       eprint = {2408.15757},
 primaryClass = {astro-ph.SR},
       adsurl = {https://ui.adsabs.harvard.edu/abs/2024ApJ...976L..21H}
}

@ARTICLE{2025A&A...695L...8R,
       author = {{Rodriguez}, Antonio C.},
        title = "{Spectroscopic detection of a 2.9-hour orbit in a long-period radio transient}",
      journal = {\aap},
         year = 2025,
        month = mar,
       volume = {695},
          eid = {L8},
        pages = {L8},
          doi = {10.1051/0004-6361/202553684},
archivePrefix = {arXiv},
       eprint = {2501.03315},
 primaryClass = {astro-ph.SR},
       adsurl = {https://ui.adsabs.harvard.edu/abs/2025A&A...695L...8R}
}

@ARTICLE{2026OJAp....967716R,
       author = {{Rodriguez}, Antonio and {El-Badry}, Kareem and {de Ruiter}, Iris and {Rajwade}, Kaustubh and {Berger}, Edo and {Connor}, Liam and {Hurley-Walker}, Natasha},
        title = "{White dwarf + M dwarf Detached Binaries in Long Period Radio Transients: Observed Binary Parameters, Evolution, and Population Constraints}",
      journal = {The Open Journal of Astrophysics},
         year = 2026,
        month = aug,
       volume = {9},
        pages = {67716},
          doi = {10.33232/001c.167716},
archivePrefix = {arXiv},
       eprint = {2604.18688},
 primaryClass = {astro-ph.SR},
       adsurl = {https://ui.adsabs.harvard.edu/abs/2026OJAp....967716R}
}

@ARTICLE{2026arXiv260628993K,
       author = {{Knigge}, Christian and {Scaringi}, Simone and {Castro Segura}, Noel and {de Martino}, Domitilla and {Veresvarska}, Martina},
        title = "{The Long-Period Radio Transient and Cataclysmic Variable ASKAP J1745-5051: Evidence for a 15,000 K White Dwarf and a Sub-Stellar Donor}",
      journal = {arXiv e-prints},
         year = 2026,
        month = jun,
          eid = {arXiv:2606.28993},
        pages = {arXiv:2606.28993},
          doi = {10.48550/arXiv.2606.28993},
archivePrefix = {arXiv},
       eprint = {2606.28993},
 primaryClass = {astro-ph.SR},
       adsurl = {https://ui.adsabs.harvard.edu/abs/2026arXiv260628993K}
}

@ARTICLE{2025NatAs...9..672D,
       author = {{de Ruiter}, I. and {Rajwade}, K.~M. and {Bassa}, C.~G. and {Rowlinson}, A. and {Wijers}, R.~A.~M.~J. and {Kilpatrick}, C.~D. and {Stefansson}, G. and {Callingham}, J.~R. and {Hessels}, J.~W.~T. and {Clarke}, T.~E. and {Peters}, W. and {Wijnands}, R.~A.~D. and {Shimwell}, T.~W. and {ter Veen}, S. and {Morello}, V. and {Zeimann}, G.~R. and {Mahadevan}, S.},
        title = "{Sporadic radio pulses from a white dwarf binary at the orbital period}",
      journal = {Nature Astronomy},
         year = 2025,
        month = may,
       volume = {9},
        pages = {672-684},
          doi = {10.1038/s41550-025-02491-0},
archivePrefix = {arXiv},
       eprint = {2408.11536},
 primaryClass = {astro-ph.HE},
       adsurl = {https://ui.adsabs.harvard.edu/abs/2025NatAs...9..672D}
}

@ARTICLE{2024MNRAS.533.2133C,
       author = {{Cooper}, A.~J. and {Wadiasingh}, Z.},
        title = "{Beyond the Rotational Deathline: Radio Emission from Ultra-long Period Magnetars}",
      journal = {\mnras},
         year = 2024,
        month = sep,
       volume = {533},
       number = {2},
        pages = {2133-2155},
          doi = {10.1093/mnras/stae1813},
archivePrefix = {arXiv},
       eprint = {2406.04135},
 primaryClass = {astro-ph.HE},
       adsurl = {https://ui.adsabs.harvard.edu/abs/2024MNRAS.533.2133C}
}

@ARTICLE{2022ApJ...934..184R,
       author = {{Ronchi}, M. and {Rea}, N. and {Graber}, V. and {Hurley-Walker}, N.},
        title = "{Long-period Pulsars as Possible Outcomes of Supernova Fallback Accretion}",
      journal = {\apj},
         year = 2022,
        month = aug,
       volume = {934},
       number = {2},
          eid = {184},
        pages = {184},
          doi = {10.3847/1538-4357/ac7cec},
archivePrefix = {arXiv},
       eprint = {2201.11704},
 primaryClass = {astro-ph.HE},
       adsurl = {https://ui.adsabs.harvard.edu/abs/2022ApJ...934..184R}
}

@ARTICLE{2026ApJ...996..141C,
       author = {{Cary}, Savannah and {Lu}, Wenbin and {Leung}, Calvin and {Wong}, Tin Long Sunny},
        title = "{Accretion from a Shock-inflated Companion: Spinning Down Neutron Stars to Hour-long Periods}",
      journal = {\apj},
         year = 2026,
        month = jan,
       volume = {996},
       number = {2},
          eid = {141},
        pages = {141},
          doi = {10.3847/1538-4357/ae1d43},
archivePrefix = {arXiv},
       eprint = {2507.10682},
 primaryClass = {astro-ph.HE},
       adsurl = {https://ui.adsabs.harvard.edu/abs/2026ApJ...996..141C}
}

@ARTICLE{2020MNRAS.496.3390B,
       author = {{Beniamini}, Paz and {Wadiasingh}, Zorawar and {Metzger}, Brian D.},
        title = "{Periodicity in recurrent fast radio bursts and the origin of ultralong period magnetars}",
      journal = {\mnras},
         year = 2020,
        month = aug,
       volume = {496},
       number = {3},
        pages = {3390-3401},
          doi = {10.1093/mnras/staa1783},
archivePrefix = {arXiv},
       eprint = {2003.12509},
 primaryClass = {astro-ph.HE},
       adsurl = {https://ui.adsabs.harvard.edu/abs/2020MNRAS.496.3390B}
}

@ARTICLE{2025ApJ...981...34Q,
       author = {{Qu}, Yuanhong and {Zhang}, Bing},
        title = "{Magnetic Interactions in White Dwarf Binaries as Mechanism for Long-period Radio Transients}",
      journal = {\apj},
         year = 2025,
        month = mar,
       volume = {981},
       number = {1},
          eid = {34},
        pages = {34},
          doi = {10.3847/1538-4357/adb1b5},
archivePrefix = {arXiv},
       eprint = {2409.05978},
 primaryClass = {astro-ph.HE},
       adsurl = {https://ui.adsabs.harvard.edu/abs/2025ApJ...981...34Q}
}

@ARTICLE{2026ApJ...999L...2Z,
       author = {{Zhong}, Yici and {Most}, Elias R.},
        title = "{Unraveling the Emission Mechanism Powering Long Period Radio Transients from Interacting White Dwarf Binaries via Kinetic Plasma Simulations}",
      journal = {\apjl},
         year = 2026,
        month = mar,
       volume = {999},
       number = {1},
          eid = {L2},
        pages = {L2},
          doi = {10.3847/2041-8213/ae4337},
archivePrefix = {arXiv},
       eprint = {2509.09057},
 primaryClass = {astro-ph.HE},
       adsurl = {https://ui.adsabs.harvard.edu/abs/2026ApJ...999L...2Z}
}

@ARTICLE{2026ApJ...997..124Y,
       author = {{Yang}, Yuan-Pei},
        title = "{Magnetic White Dwarf─M Dwarf Binaries in Pre-mCV Phase as Special Population of Long-period Radio Transients}",
      journal = {\apj},
         year = 2026,
        month = jan,
       volume = {997},
       number = {1},
          eid = {124},
        pages = {124},
          doi = {10.3847/1538-4357/ae2864},
archivePrefix = {arXiv},
       eprint = {2509.09224},
 primaryClass = {astro-ph.HE},
       adsurl = {https://ui.adsabs.harvard.edu/abs/2026ApJ...997..124Y}
}

@ARTICLE{2013PASP..125..306F,
       author = {{Foreman-Mackey}, Daniel and {Hogg}, David W. and {Lang}, Dustin and {Goodman}, Jonathan},
        title = "{emcee: The MCMC Hammer}",
      journal = {\pasp},
         year = 2013,
        month = mar,
       volume = {125},
       number = {925},
        pages = {306},
          doi = {10.1086/670067},
archivePrefix = {arXiv},
       eprint = {1202.3665},
 primaryClass = {astro-ph.IM},
       adsurl = {https://ui.adsabs.harvard.edu/abs/2013PASP..125..306F}
}

@ARTICLE{1977ApJ...212L.121C,
       author = {{Cowley}, A.~P. and {Crampton}, D.},
        title = "{A preliminary model for the X-ray binary AM Herculis.}",
      journal = {\apjl},
         year = 1977,
        month = mar,
       volume = {212},
        pages = {L121-L124},
          doi = {10.1086/182389},
       adsurl = {https://ui.adsabs.harvard.edu/abs/1977ApJ...212L.121C}
}

@ARTICLE{2011ApJS..194...28K,
       author = {{Knigge}, Christian and {Baraffe}, Isabelle and {Patterson}, Joseph},
        title = "{The Evolution of Cataclysmic Variables as Revealed by Their Donor Stars}",
      journal = {\apjs},
         year = 2011,
        month = jun,
       volume = {194},
       number = {2},
          eid = {28},
        pages = {28},
          doi = {10.1088/0067-0049/194/2/28},
archivePrefix = {arXiv},
       eprint = {1102.2440},
 primaryClass = {astro-ph.SR},
       adsurl = {https://ui.adsabs.harvard.edu/abs/2011ApJS..194...28K}
}

@INPROCEEDINGS{1985ASSL..113..151L,
       author = {{Liebert}, J. and {Stockman}, H.~S.},
        title = "{The AM Herculis Magnetic Variables}",
    booktitle = {Cataclysmic Variables and Low-Mass X-ray Binaries},
         year = 1985,
       editor = {{Lamb}, D.~Q. and {Patterson}, J.},
        month = jan,
        pages = {151},
          doi = {10.1007/978-94-009-5319-2_20},
       adsurl = {https://ui.adsabs.harvard.edu/abs/1985ASSL..113..151L}
}

@ARTICLE{2024MNRAS.528..676G,
       author = {{Galiullin}, Ilkham and {Rodriguez}, Antonio C. and {Kulkarni}, Shrinivas R. and {Sunyaev}, Rashid and {Gilfanov}, Marat and {Bikmaev}, Ilfan and {Yungelson}, Lev and {van Roestel}, Jan and {G{\"a}nsicke}, Boris T. and {Khamitov}, Irek and {Szkody}, Paula and {El-Badry}, Kareem and {Suslikov}, Mikhail and {Prince}, Thomas A. and {Buntov}, Mikhail and {Caiazzo}, Ilaria and {Gorbachev}, Mark and {Graham}, Matthew J. and {Gumerov}, Rustam and {Irtuganov}, Eldar and {Laher}, Russ R. and {Medvedev}, Pavel and {Riddle}, Reed and {Rusholme}, Ben and {Sakhibullin}, Nail and {Sklyanov}, Alexander and {Vanderbosch}, Zachary P.},
        title = "{A joint SRG/eROSITA + ZTF search: Discovery of a 97-min period eclipsing cataclysmic variable with evidence of a brown dwarf secondary}",
      journal = {\mnras},
         year = 2024,
        month = feb,
       volume = {528},
       number = {1},
        pages = {676-692},
          doi = {10.1093/mnras/stae012},
archivePrefix = {arXiv},
       eprint = {2401.04178},
 primaryClass = {astro-ph.HE},
       adsurl = {https://ui.adsabs.harvard.edu/abs/2024MNRAS.528..676G}
}

@ARTICLE{2025PASP..137a4201R,
       author = {{Rodriguez}, Antonio C. and {El-Badry}, Kareem and {Suleimanov}, Valery and {Pala}, Anna F. and {Kulkarni}, Shrinivas R. and {Gaensicke}, Boris and {Mori}, Kaya and {Rich}, R. Michael and {Sarkar}, Arnab and {Bao}, Tong and {Lopes de Oliveira}, Raimundo and {Ramsay}, Gavin and {Szkody}, Paula and {Graham}, Matthew and {Prince}, Thomas A. and {Caiazzo}, Ilaria and {Vanderbosch}, Zachary P. and {van Roestel}, Jan and {Das}, Kaustav K. and {Qin}, Yu-Jing and {Kasliwal}, Mansi M. and {Wold}, Avery and {Groom}, Steven L. and {Reiley}, Daniel and {Riddle}, Reed},
        title = "{Cataclysmic Variables and AM CVn Binaries in SRG/eROSITA + Gaia: Volume Limited Samples, X-Ray Luminosity Functions, and Space Densities}",
      journal = {\pasp},
         year = 2025,
        month = jan,
       volume = {137},
       number = {1},
          eid = {014201},
        pages = {014201},
          doi = {10.1088/1538-3873/ada185},
archivePrefix = {arXiv},
       eprint = {2408.16053},
 primaryClass = {astro-ph.HE},
       adsurl = {https://ui.adsabs.harvard.edu/abs/2025PASP..137a4201R}
}

@ARTICLE{2004AJ....127.3516G,
       author = {{Golimowski}, D.~A. and {Leggett}, S.~K. and {Marley}, M.~S. and {Fan}, X. and {Geballe}, T.~R. and {Knapp}, G.~R. and {Vrba}, F.~J. and {Henden}, A.~A. and {Luginbuhl}, C.~B. and {Guetter}, H.~H. and {Munn}, J.~A. and {Canzian}, B. and {Zheng}, W. and {Tsvetanov}, Z.~I. and {Chiu}, K. and {Glazebrook}, K. and {Hoversten}, E.~A. and {Schneider}, D.~P. and {Brinkmann}, J.},
        title = "{L' and M' Photometry of Ultracool Dwarfs}",
      journal = {\aj},
         year = 2004,
        month = jun,
       volume = {127},
       number = {6},
        pages = {3516-3536},
          doi = {10.1086/420709},
archivePrefix = {arXiv},
       eprint = {astro-ph/0402475},
 primaryClass = {astro-ph},
       adsurl = {https://ui.adsabs.harvard.edu/abs/2004AJ....127.3516G}
}

@ARTICLE{1983ApJS...53..523W,
       author = {{Williams}, G.},
        title = "{Spectroscopy of cataclysmic variables. I. Observations.}",
      journal = {\apjs},
         year = 1983,
        month = nov,
       volume = {53},
        pages = {523-552},
          doi = {10.1086/190900},
       adsurl = {https://ui.adsabs.harvard.edu/abs/1983ApJS...53..523W}
}

@ARTICLE{1988MNRAS.233..513E,
       author = {{Echevarria}, J.},
        title = "{A statistical analysis of the emission line ratios in cataclysmic variables.}",
      journal = {\mnras},
         year = 1988,
        month = aug,
       volume = {233},
        pages = {513-527},
          doi = {10.1093/mnras/233.3.513},
       adsurl = {https://ui.adsabs.harvard.edu/abs/1988MNRAS.233..513E}
}

@ARTICLE{2026A&A...712A.116H,
       author = {{Hern{\'a}ndez-D{\'\i}az}, Santiago and {Stelzer}, Beate and {Mu{\~n}oz-Giraldo}, Daniela},
        title = "{Balmer decrements as a new diagnostic for period-bounce cataclysmic variable stars}",
      journal = {\aap},
         year = 2026,
        month = aug,
       volume = {712},
          eid = {A116},
        pages = {A116},
          doi = {10.1051/0004-6361/202659008},
archivePrefix = {arXiv},
       eprint = {2606.09627},
 primaryClass = {astro-ph.SR},
       adsurl = {https://ui.adsabs.harvard.edu/abs/2026A&A...712A.116H}
}

@ARTICLE{1991AJ....101.1929W,
       author = {{Williams}, Glen A.},
        title = "{Hydrogen line emission from accretion disks. II - A grid of models with applications to the optically thin regions of disks.}",
      journal = {\aj},
         year = 1991,
        month = may,
       volume = {101},
        pages = {1929-1941},
          doi = {10.1086/115818},
       adsurl = {https://ui.adsabs.harvard.edu/abs/1991AJ....101.1929W}
}

@ARTICLE{2000MNRAS.318..440M,
       author = {{Mason}, Elena and {Skidmore}, Warren and {Howell}, Steve B. and {Ciardi}, David R. and {Littlefair}, Stuart and {Dhillon}, V.~S.},
        title = "{Investigating the structure of the accretion disc in WZ Sge from multiwaveband time-resolved spectroscopic observations - II}",
      journal = {\mnras},
         year = 2000,
        month = oct,
       volume = {318},
       number = {2},
        pages = {440-452},
          doi = {10.1046/j.1365-8711.2000.03780.x},
archivePrefix = {arXiv},
       eprint = {astro-ph/0006085},
 primaryClass = {astro-ph},
       adsurl = {https://ui.adsabs.harvard.edu/abs/2000MNRAS.318..440M}
}

@INPROCEEDINGS{1998ASPC..137..446W,
       author = {{Wheatley}, P.~J. and {Ramsay}, G.},
        title = "{EF ERI in a low accretion state}",
    booktitle = {Wild Stars in the Old West},
         year = 1998,
       editor = {{Howell}, S. and {Kuulkers}, E. and {Woodward}, C.},
       series = {Astronomical Society of the Pacific Conference Series},
       volume = {137},
        month = jan,
        pages = {446},
       adsurl = {https://ui.adsabs.harvard.edu/abs/1998ASPC..137..446W}
}

@ARTICLE{2006ApJ...652..709H,
       author = {{Howell}, Steve B. and {Walter}, Frederick M. and {Harrison}, Thomas E. and {Huber}, Mark E. and {Becker}, Robert H. and {White}, Richard L.},
        title = "{Mass Determination and Detection of the Onset of Chromospheric Activity for the Substellar Object in EF Eridani}",
      journal = {\apj},
         year = 2006,
        month = nov,
       volume = {652},
       number = {1},
        pages = {709-723},
          doi = {10.1086/507603},
archivePrefix = {arXiv},
       eprint = {astro-ph/0607140},
 primaryClass = {astro-ph},
       adsurl = {https://ui.adsabs.harvard.edu/abs/2006ApJ...652..709H}
}

@ARTICLE{2000A&A...354L..49B,
       author = {{Beuermann}, K. and {Wheatley}, P. and {Ramsay}, G. and {Euchner}, F. and {G{\"a}nsicke}, B.~T.},
        title = "{Evidence for a substellar secondary in the magnetic cataclysmic binary EF Eridani}",
      journal = {\aap},
         year = 2000,
        month = feb,
       volume = {354},
        pages = {L49-L52},
          doi = {10.48550/arXiv.astro-ph/0001183},
archivePrefix = {arXiv},
       eprint = {astro-ph/0001183},
 primaryClass = {astro-ph},
       adsurl = {https://ui.adsabs.harvard.edu/abs/2000A&A...354L..49B}
}

@ARTICLE{2025MNRAS.544..309K,
       author = {{Khangale}, Z.~N. and {Potter}, S.~B. and {Buckley}, D.~A.~H. and {Barrett}, P.~E.},
        title = "{Phase-resolved spectroscopic observations of the magnetic cataclysmic binary EF Eridani: revealing complex magnetic accretion during a high state}",
      journal = {\mnras},
         year = 2025,
        month = nov,
       volume = {544},
       number = {1},
        pages = {309-320},
          doi = {10.1093/mnras/staf1754},
archivePrefix = {arXiv},
       eprint = {2510.08266},
 primaryClass = {astro-ph.SR},
       adsurl = {https://ui.adsabs.harvard.edu/abs/2025MNRAS.544..309K}
}

@ARTICLE{2026A&A...710L..27I,
       author = {{Imbrogno}, M. and {Veresvarska}, M. and {Wang}, Y.~L. and {Rea}, N. and {Coti Zelati}, F. and {Rose}, K. and {Pritchard}, J. and {de Martino}, D. and {Scaringi}, S. and {Wang}, Z. and {Kaplan}, D.~L.},
        title = "{The X-ray emission of the long-period transient and accreting cataclysmic variable ASKAP J174508.9-505149}",
      journal = {\aap},
         year = 2026,
        month = jun,
       volume = {710},
          eid = {L27},
        pages = {L27},
          doi = {10.1051/0004-6361/202660831},
archivePrefix = {arXiv},
       eprint = {2606.05842},
 primaryClass = {astro-ph.HE},
       adsurl = {https://ui.adsabs.harvard.edu/abs/2026A&A...710L..27I}
}

@ARTICLE{2026NatAs..10..522H,
       author = {{Horv{\'a}th}, Csan{\'a}d and {Rea}, Nanda and {Hurley-Walker}, Natasha and {McSweeney}, Samuel J. and {Perley}, Richard A. and {Lenc}, Emil},
        title = "{A binary model of long-period radio transients and white dwarf pulsars}",
      journal = {Nature Astronomy},
         year = 2026,
        month = apr,
       volume = {10},
        pages = {522-530},
          doi = {10.1038/s41550-025-02760-y},
archivePrefix = {arXiv},
       eprint = {2507.15352},
 primaryClass = {astro-ph.HE},
       adsurl = {https://ui.adsabs.harvard.edu/abs/2026NatAs..10..522H}
}

@ARTICLE{2023Natur.619..487H,
       author = {{Hurley-Walker}, N. and {Rea}, N. and {McSweeney}, S.~J. and {Meyers}, B.~W. and {Lenc}, E. and {Heywood}, I. and {Hyman}, S.~D. and {Men}, Y.~P. and {Clarke}, T.~E. and {Coti Zelati}, F. and {Price}, D.~C. and {Horv{\'a}th}, C. and {Galvin}, T.~J. and {Anderson}, G.~E. and {Bahramian}, A. and {Barr}, E.~D. and {Bhat}, N.~D.~R. and {Caleb}, M. and {Dall'Ora}, M. and {de Martino}, D. and {Giacintucci}, S. and {Morgan}, J.~S. and {Rajwade}, K.~M. and {Stappers}, B. and {Williams}, A.},
        title = "{A long-period radio transient active for three decades}",
      journal = {\nat},
         year = 2023,
        month = jul,
       volume = {619},
       number = {7970},
        pages = {487-490},
          doi = {10.1038/s41586-023-06202-5},
archivePrefix = {arXiv},
       eprint = {2503.08036},
 primaryClass = {astro-ph.HE},
       adsurl = {https://ui.adsabs.harvard.edu/abs/2023Natur.619..487H}
}

@Article{         harris2020array,
 title         = {Array programming with {NumPy}},
 author        = {Charles R. Harris and K. Jarrod Millman and St{\'{e}}fan J.
                 van der Walt and Ralf Gommers and Pauli Virtanen and David
                 Cournapeau and Eric Wieser and Julian Taylor and Sebastian
                 Berg and Nathaniel J. Smith and Robert Kern and Matti Picus
                 and Stephan Hoyer and Marten H. van Kerkwijk and Matthew
                 Brett and Allan Haldane and Jaime Fern{\'{a}}ndez del
                 R{\'{i}}o and Mark Wiebe and Pearu Peterson and Pierre
                 G{\'{e}}rard-Marchant and Kevin Sheppard and Tyler Reddy and
                 Warren Weckesser and Hameer Abbasi and Christoph Gohlke and
                 Travis E. Oliphant},
 year          = {2020},
 month         = sep,
 journal       = {Nature},
 volume        = {585},
 number        = {7825},
 pages         = {357--362},
 doi           = {10.1038/s41586-020-2649-2},
 publisher     = {Springer Science and Business Media {LLC}},
 url           = {https://doi.org/10.1038/s41586-020-2649-2}
}

@misc{Hunter:2007,
  Author    = {Hunter, J. D.},
  Title     = {Matplotlib: A 2D graphics environment},
  Journal   = {Computing in Science \& Engineering},
  Volume    = {9},
  Number    = {3},
  Pages     = {90--95},
  publisher = {IEEE COMPUTER SOC},
  doi       = {10.1109/MCSE.2007.55},
  year      = 2007
}

@misc{mckinney2010data,
  title={Data structures for statistical computing in python},
  author={McKinney, Wes},
  publisher={Proceedings of the 9th Python in Science Conference},
  volume={445},
  pages={51--56},
  year={2010},
  organization={Austin, TX}
}

@misc{astropy:2013,
   title={Astropy: A community Python package for astronomy},
   volume={558},
   ISSN={1432-0746},
   url={http://dx.doi.org/10.1051/0004-6361/201322068},
   DOI={10.1051/0004-6361/201322068},
   journal={Astronomy & Astrophysics},
   publisher={EDP Sciences},
   author={Robitaille, Thomas P. and Tollerud, Erik J. and Greenfield, Perry and Droettboom, Michael and Bray, Erik and Aldcroft, Tom and Davis, Matt and Ginsburg, Adam and Price-Whelan, Adrian M. and et al.},
   year={2013},
   month={Sep},
   pages={A33}}

@ARTICLE{2020SciPy-NMeth,
  author  = {Virtanen, Pauli and Gommers, Ralf and Oliphant, Travis E. and
            Haberland, Matt and Reddy, Tyler and Cournapeau, David and
            Burovski, Evgeni and Peterson, Pearu and Weckesser, Warren and
            Bright, Jonathan and {van der Walt}, St{\'e}fan J. and
            Brett, Matthew and Wilson, Joshua and Millman, K. Jarrod and
            Mayorov, Nikolay and Nelson, Andrew R. J. and Jones, Eric and
            Kern, Robert and Larson, Eric and Carey, C J and
            Polat, {\.I}lhan and Feng, Yu and Moore, Eric W. and
            {VanderPlas}, Jake and Laxalde, Denis and Perktold, Josef and
            Cimrman, Robert and Henriksen, Ian and Quintero, E. A. and
            Harris, Charles R. and Archibald, Anne M. and
            Ribeiro, Ant{\^o}nio H. and Pedregosa, Fabian and
            {van Mulbregt}, Paul and {SciPy 1.0 Contributors}},
  title   = {{{SciPy} 1.0: Fundamental Algorithms for Scientific
            Computing in Python}},
  journal = {Nature Methods},
  year    = {2020},
  volume  = {17},
  pages   = {261--272},
  adsurl  = {https://rdcu.be/b08Wh},
  doi     = {10.1038/s41592-019-0686-2},
}

@software{2021zndo....598352N,
       author = {{Newville}, Matthew and {Otten}, Renee and {Nelson}, Andrew and {Stensitzki}, Till and {Ingargiola}, Antonino and {Allan}, Daniel and {Fox}, Austin and {Carter}, Faustin and {Rawlik}, Michal},
        title = "{LMFIT: Non-Linear Least-Squares Minimization and Curve-Fitting for Python}",
         year = 2025,
        month = jul,
          eid = {10.5281/zenodo.598352},
          doi = {10.5281/zenodo.598352},
      version = {1.3.4},
    publisher = {Zenodo},
       adsurl = {https://ui.adsabs.harvard.edu/abs/2021zndo....598352N}
}

@ARTICLE{2005Natur.434...50H,
       author = {{Hyman}, Scott D. and {Lazio}, T. Joseph W. and {Kassim}, Namir E. and {Ray}, Paul S. and {Markwardt}, Craig B. and {Yusef-Zadeh}, Farhad},
        title = "{A powerful bursting radio source towards the Galactic Centre}",
      journal = {\nat},
         year = 2005,
        month = mar,
       volume = {434},
       number = {7029},
        pages = {50-52},
          doi = {10.1038/nature03400},
archivePrefix = {arXiv},
       eprint = {astro-ph/0503052},
 primaryClass = {astro-ph},
       adsurl = {https://ui.adsabs.harvard.edu/abs/2005Natur.434...50H}
}

@ARTICLE{2022Natur.601..526H,
       author = {{Hurley-Walker}, N. and {Zhang}, X. and {Bahramian}, A. and {McSweeney}, S.~J. and {O'Doherty}, T.~N. and {Hancock}, P.~J. and {Morgan}, J.~S. and {Anderson}, G.~E. and {Heald}, G.~H. and {Galvin}, T.~J.},
        title = "{A radio transient with unusually slow periodic emission}",
      journal = {\nat},
         year = 2022,
        month = jan,
       volume = {601},
       number = {7894},
        pages = {526-530},
          doi = {10.1038/s41586-021-04272-x},
archivePrefix = {arXiv},
       eprint = {2503.08033},
 primaryClass = {astro-ph.HE},
       adsurl = {https://ui.adsabs.harvard.edu/abs/2022Natur.601..526H}
}

@ARTICLE{2024NatAs...8.1159C,
       author = {{Caleb}, M. and {Lenc}, E. and {Kaplan}, D.~L. and {Murphy}, T. and {Men}, Y.~P. and {Shannon}, R.~M. and {Ferrario}, L. and {Rajwade}, K.~M. and {Clarke}, T.~E. and {Giacintucci}, S. and {Hurley-Walker}, N. and {Hyman}, S.~D. and {Lower}, M.~E. and {McSweeney}, Sam and {Ravi}, V. and {Barr}, E.~D. and {Buchner}, S. and {Flynn}, C.~M.~L. and {Hessels}, J.~W.~T. and {Kramer}, M. and {Pritchard}, J. and {Stappers}, B.~W.},
        title = "{An emission-state-switching radio transient with a 54-minute period}",
      journal = {Nature Astronomy},
         year = 2024,
        month = sep,
       volume = {8},
        pages = {1159-1168},
          doi = {10.1038/s41550-024-02277-w},
archivePrefix = {arXiv},
       eprint = {2407.12266},
 primaryClass = {astro-ph.HE},
       adsurl = {https://ui.adsabs.harvard.edu/abs/2024NatAs...8.1159C}
}

@ARTICLE{dragons,
       author = {{Labrie}, K. and {Simpson}, C. and {Cardenes}, R. and {Turner}, J. and {Soraisam}, M. and {Quint}, B. and {Oberdorf}, O. and {Placco}, V.~M. and {Berke}, D. and {Smirnova}, O. and {Conseil}, S. and {Vacca}, W.~D. and {Thomas-Osip}, J.},
        title = "{DRAGONS-A Quick Overview}",
      journal = {Research Notes of the American Astronomical Society},
         year = 2023,
        month = oct,
       volume = {7},
       number = {10},
          eid = {214},
        pages = {214},
          doi = {10.3847/2515-5172/ad0044},
archivePrefix = {arXiv},
       eprint = {2310.03048},
 primaryClass = {astro-ph.IM},
       adsurl = {https://ui.adsabs.harvard.edu/abs/2023RNAAS...7..214L}
}

@ARTICLE{decaps,
       author = {{Saydjari}, Andrew K. and {Schlafly}, Edward F. and {Lang}, Dustin and {Meisner}, Aaron M. and {Green}, Gregory M. and {Zucker}, Catherine and {Zelko}, Ioana and {Speagle}, Joshua S. and {Daylan}, Tansu and {Lee}, Albert and {Valdes}, Francisco and {Schlegel}, David and {Finkbeiner}, Douglas P.},
        title = "{The Dark Energy Camera Plane Survey 2 (DECaPS2): More Sky, Less Bias, and Better Uncertainties}",
      journal = {\apjs},
         year = 2023,
        month = feb,
       volume = {264},
       number = {2},
          eid = {28},
        pages = {28},
          doi = {10.3847/1538-4365/aca594},
archivePrefix = {arXiv},
       eprint = {2206.11909},
 primaryClass = {astro-ph.GA},
       adsurl = {https://ui.adsabs.harvard.edu/abs/2023ApJS..264...28S}
}

@software{2026zndo..19636730B,
       author = {{Bradley}, Larry and {Sip{\H{o}}cz}, Brigitta M. and {Robitaille}, T.~P. and {Tollerud}, E.~J. and {Vin{\'\i}cius}, Z{\'e} and {Deil}, Christoph and {Barbary}, Kyle and {Wilson}, Tom J. and {Busko}, Ivo and {Donath}, Axel and {G{\"u}nther}, Hans Moritz and {Cara}, Mihai and {Lim}, P.~L. and {Me{\ss}linger}, Sebastian and {Conseil}, Simon and {Droettboom}, Michael and {Bostroem}, K. Azalee and {Bray}, E.~M. and {Andersen Bratholm}, Lars and {Burnett}, Zach and {Jamieson}, William and {Ginsburg}, Adam and {Taranu}, Dan and {Barentsen}, Geert and {Craig}, Matthew W. and {Morris}, Brett M. and {Perrin}, Marshall and {Rathi}, Shivangee},
        title = "{Photutils}",
         year = 2026,
        month = apr,
          eid = {10.5281/zenodo.19636730},
          doi = {10.5281/zenodo.19636730},
      version = {3.0.0},
    publisher = {Zenodo},
       adsurl = {https://ui.adsabs.harvard.edu/abs/2026zndo..19636730B}
}

@ARTICLE{1987PASP...99..191S,
       author = {{Stetson}, Peter B.},
        title = "{DAOPHOT: A Computer Program for Crowded-Field Stellar Photometry}",
      journal = {\pasp},
         year = 1987,
        month = mar,
       volume = {99},
        pages = {191},
          doi = {10.1086/131977},
       adsurl = {https://ui.adsabs.harvard.edu/abs/1987PASP...99..191S}
}

@ARTICLE{1980ApJS...42..351D,
       author = {{Drake}, S.~A. and {Ulrich}, R.~K.},
        title = "{The emission-line spectrum from a slab of hydrogen at moderate to high densities.}",
      journal = {\apjs},
         year = 1980,
        month = feb,
       volume = {42},
        pages = {351-383},
          doi = {10.1086/190654},
       adsurl = {https://ui.adsabs.harvard.edu/abs/1980ApJS...42..351D}
}

@ARTICLE{2020A&A...642A.134S,
       author = {{Schwope}, A.~D. and {Worpel}, H. and {Traulsen}, I. and {Sablowski}, D.},
        title = "{The various accretion modes of AM Herculis: Clues from multi-wavelength observations in high accretion states}",
      journal = {\aap},
         year = 2020,
        month = oct,
       volume = {642},
          eid = {A134},
        pages = {A134},
          doi = {10.1051/0004-6361/202037714},
archivePrefix = {arXiv},
       eprint = {2008.08402},
 primaryClass = {astro-ph.SR},
       adsurl = {https://ui.adsabs.harvard.edu/abs/2020A&A...642A.134S}
}

@ARTICLE{1982ApJ...257..686S,
       author = {{Szkody}, P. and {Raymond}, J.~C. and {Capps}, R.~W.},
        title = "{The low state of AM HER : observations from 0.12 to 10 microns.}",
      journal = {\apj},
         year = 1982,
        month = jun,
       volume = {257},
        pages = {686-694},
          doi = {10.1086/160023},
       adsurl = {https://ui.adsabs.harvard.edu/abs/1982ApJ...257..686S}
}

@ARTICLE{2008ApJ...683..409C,
       author = {{Campbell}, Ryan K. and {Harrison}, Thomas E. and {Kafka}, Stella},
        title = "{Cyclotron Modeling Phase-Resolved Infrared Spectroscopy of Polars. III. AM Herculis and ST Leo Minoris}",
      journal = {\apj},
         year = 2008,
        month = aug,
       volume = {683},
       number = {1},
        pages = {409-423},
          doi = {10.1086/589179},
       adsurl = {https://ui.adsabs.harvard.edu/abs/2008ApJ...683..409C}
}

@ARTICLE{vizier,
  author  = {{Ochsenbein}, F. and {Bauer}, P. and {Marcout}, J.},
  title   = "{The VizieR database of astronomical catalogues}",
  journal = {\aaps},
  year    = {2000},
  volume  = {143},
  pages   = {23-32},
  doi     = {10.1051/aas:2000169}
}







\appendix
\section{Notes on the radio period}\label{app:timing}
Using a set of 18 times of arrival (TOAs), \citet{anumarlapudi_j1448} proposed a timing period of 93.8513(8)\,min. 
However, these TOAs come from two sets of observations separated by a year (see Table~B1 in \citealt{anumarlapudi_j1448}). 
In addition, the pulse shapes of \jf\ are extremely diverse, which means that extra noise terms need to be added to account for this variability. 
This combination of ill-sampled and less precisely defined TOAs means that the timing solution proposed by \citet{anumarlapudi_j1448} is a plausible solution, but may not be the unique one that determines the system. 
Additional TOAs are required to robustly constrain the orbital parameters, whilst accounting for the pulse variability. 
In addition, while the radio period is constrained to sub-minute, the error on the predicted phase (while phase-folding optical and radio data) is proportional to the error on the individual TOAs, which is currently $\approx$ 10\,min. 
This means that i) the radio period can be subject to change with additional data, and ii) phase folding the radio TOAs to the \ac{rv} curve will not lead to meaningful constraints.


\bsp	
\label{lastpage}
\end{document}